\documentclass[preprint,showpacs,preprintnumbers,amsmath,amssymb]{revtex4}

\usepackage{graphicx}
\usepackage{dcolumn}
\usepackage{bm}

\begin{document}

\title{
Quantum effect in the diffusion along a potential barrier: 
Comments on the synthesis of superheavy elements}
\author{Noboru Takigawa$^1$}
\email{takigawa@nucl.phys.tohoku.ac.jp}
\author{Sakir Ayik$^2$}
\email{ayik@tntech.edu}
\author{Kouhei Washiyama$^1$}
\email{washi@nucl.phys.tohoku.ac.jp}
\author{Sachie Kimura$^3$}
\email{kimura@lns.infn.it}
\affiliation{$^1$Department of Physics, Tohoku University, 
Sendai 980-8578, Japan \\
$^2$Physics Department,
Tennessee Technological University, Cookeville, TN 38505, USA\\
$^3$Laboratorio Nazionale del Sud, INFN, via Santa Sofia, 62, 95123
Catania, Italy
}

\date{\today}

\begin{abstract}

We discuss a quantum effect in the diffusion process 
by developing a theory, which takes 
the finite curvature 
of the potential field into account. 
The transport coefficients of our theory 
satisfy the well-known fluctuation-dissipation theorem 
in the limit of Markovian approximation 
in the cases of diffusion in a flat potential and in a potential well.
For the diffusion along a potential barrier, 
the diffusion coefficient can be related to the 
friction coefficient 
by an analytic continuation of the fluctuation-dissipation 
theorem for the case of diffusion along a potential well in the 
asymptotic time, but contains strong non-Markovian effects at short times. 
By applying our theory to the case of realistic values of the 
temperature, the barrier curvature, and the friction coefficient, we show that 
the quantum effects will play significant roles 
in describing the synthesis of superheavy elements, i.e., the 
evolution from the fusion barrier to the conditional saddle, 
in terms of a diffusion process. 
We especially point out the importance of the memory effect, which 
increases at lower temperatures. It makes the net quantum effects 
enhance the probability of crossing the conditional saddle.

\end{abstract}


\maketitle

\section{Introduction}

Diffusion processes take place in a variety of problems
\cite{kubo,hangi90,weiss,llsp,ar94,hh97}. The 
concept has recently been applied to theoretically describing the synthesis 
of superheavy elements \cite{fddabe,ohta}. It is now well accepted that 
it is not sufficient for the two nuclei in heavy-ion collisions 
to overcome the fusion (i.e., the Coulomb) barrier 
in order to form a heavy compound nucleus such as superheavy elements. 
Since the conditional saddle is located inside the fusion barrier for 
heavy-ion collisions between two heavy nuclei, 
two nuclei have to further progress inwards to approach inside the 
conditional saddle. The idea of Refs.~\cite{fddabe,ohta}, 
called fluctuation-dissipation dynamics, is to describe the time 
evolution from the fusion barrier to the 
conditional saddle as a diffusion process. 

Though this approach is very attractive and is offering much  
useful information, one needs to examine the applicability of one of the 
basic assumptions made so far, i.e., the use of the standard 
fluctuation-dissipation theorem which holds at high temperatures 
to relate the diffusion coefficients 
to the friction coefficients. Since superheavy elements are stabilized 
by shell correction energies, one has to 
synthesize them at reasonably low energies, as low as 1 MeV or below. 
On the other hand, the curvature of the conditional saddle 
is also of the order 
of 1 MeV. It is thus required to carefully study quantum effects. 
An interesting issue is to explore the connection between the 
diffusion and friction coefficients in the diffusion process 
along a potential barrier under such circumstances. 
Though the generalization of the Einstein 
relation to the case of diffusion along a potential well is 
well known, the modification in the case of a diffusion along 
a potential barrier has so far been discussed only in a limited number of 
papers \cite{ar94,hh97,rh03}. 

The aim of this paper is to examine this quantum effect caused by the 
finite curvature of the potential barrier and by 
the low temperature aspect of the diffusion process. 
To this end, we first develop a novel quantum diffusion theory 
which leads to a Fokker-Planck equation with non-Markovian transport 
coefficients. We then apply it to the situation relevant to 
the synthesis of superheavy elements.
We will show that the quantum effects, especially the non-Markovian 
effects, are very important.

In Sec. II, we briefly sketch the derivation of 
the Fokker-Planck equation with non-Markovian 
transport coefficients, which include quantum effects. 
In Sec. III we discuss the quantum effects on the diffusion 
coefficient.
In Sec. IV we apply the formalism to analyze the quantum effects 
by assuming the parameters which are relevant to describe the 
diffusion process in the synthesis of superheavy elements.
We summarize the paper in Sec. V.

\section{Quantum diffusion equation with non-Markovian transport 
coefficients}

We derive the required diffusion equation by extending  
the quasilinear response theory 
developed in Ref. \cite{tak81} so as to include the 
curvature of the potential barrier.  
Details of the derivation will be published elsewhere \cite{tak02}.
Here we sketch the main steps:
(1) We start form the von Neumann equation for the system 
consisting of space A for macroscopic degrees of freedom and 
space B, for microscopic degrees of freedom. (2) We introduce 
the classical trajectory given by $(q(t)$, $p(t))$, $t$ being the time and  
${\hat q}$ and ${\hat p}$ being the coordinate and the conjugate 
momentum operators of the macroscopic degrees of freedom. (3) We move to 
the Galilei transformed coordinate system 
specified by $q(t)$ and $p(t)$. (4) We keep 
only up to the second order terms of the potential and the coupling 
Hamiltonian in the expansion with respect to the fluctuations around the 
classical trajectory. (5) We solve the coupled equations describing the 
A and B spaces in the lowest order approximation concerning the 
fluctuating force. (6) We make a Wigner transform of the resultant extended 
von Neumann equation for subspace A. 

Denoting the Wigner transform of the 
density operator of the subspace A by $D_{AW}$, we finally obtain
\begin{eqnarray}
&&\frac{\partial}{\partial t} D_{AW}(p,q,t)
=\Biggl(-\frac{1}{M}p_\alpha \frac{\partial }{\partial q_\alpha}
+Cq_\alpha\frac{\partial}{\partial p_\alpha}\nonumber\\
&&{}-\chi^{(-E)}_{\alpha \beta} q_\beta \frac{\partial }{\partial p_\alpha}
+\chi^{(-O)}_{\alpha \beta}
\frac{\partial }{\partial p_\alpha} p_\beta
+\chi^{(+O)}_{\alpha \beta} \frac{\partial^2}{\partial p_\alpha 
\partial q_\beta} \nonumber \\
&&{}+\chi^{(+E)}_{\alpha \beta}
\frac{\partial^2}{\partial p_\alpha \partial p_\beta}\Biggr) 
D_{AW}(p,q,t).
\label{QFPE} 
\end{eqnarray}
The non-Markovian property of the diffusion process 
due to quantum effects, more specifically the effects of the 
curvature of the potential barrier, 
is hidden in the transport coefficients, 
which are generalized from the 1st and 0th moments of the 
response and correlation functions, $\chi^{(-)}$ and  $\chi^{(+)}$,
e.g. as
\begin{eqnarray}
\chi^{(-O)}_{\alpha \beta}(t)&=&\int^t_{t_0}dt_1 {\cal S}(t,t_1)
\chi^{(-)}_{\alpha \beta}(t,t_1),
\label{ROmoment} \\
\chi^{(+E)}_{\alpha \beta}(t)&=&\int^t_{t_0}dt_1 {\cal C}(t,t_1)
\chi^{(+)}_{\alpha \beta}(t,t_1).
\label{CEmoment} 
\end{eqnarray}
The ${\cal C}$ and ${\cal S}$ are defined by using the curvature of the 
potential $C$ as
\begin{eqnarray}
{\cal C}(t,t_1)&=&\cos[\Omega(t-t_1)] ,
\label{mfuncCP} \\
{\cal S}(t,t_1)&=&\frac{1}{{\sqrt {MC}}}\sin[\Omega(t-t_1)] ,
\label{mfuncSP}
\end{eqnarray}
with $\Omega={\sqrt{C/M}}$ 
when $C\geq0$, i.e., for the motion in a potential well, and 
\begin{eqnarray}
{\cal C}(t,t_1)&=&\cosh[\Omega(t-t_1)] ,
\label{mfuncCN} \\
{\cal S}(t,t_1)&=&\frac{1}{{\sqrt {M \vert C \vert}}}\sinh[\Omega(t-t_1)] ,
\label{mfuncSN}
\end{eqnarray}
with $\Omega={\sqrt{\vert C \vert/M}}$ 
when $C<0$, i.e. for a motion along a potential barrier.
The response and the correlation functions, 
$\chi^{(-)}_{\alpha \beta}(t,t_1)$ and $\chi^{(+)}_{\alpha \beta}(t,t_1)$, 
are given by 
\begin{eqnarray}
\chi^{(-)}_{\alpha \beta}(t,t_1)&=&\frac{i}{\hbar}
Tr_B([{\hat f}_\alpha(t),{\hat f}_\beta(t_1)]{\hat D}_B(t_1)),
\label{rfunct} \\
\chi^{(+)}_{\alpha \beta}(t,t_1)&=&\frac{1}{2}
Tr_B([{\hat f}_\alpha(t),{\hat f}_\beta(t_1)]_+{\hat D}_B(t_1)),
\label{cfunct}
\end{eqnarray}
in terms of the fluctuation force ${\hat f}_\alpha(t)$ defined by 
\begin{eqnarray}
{\hat f}_\alpha(t)={\hat u}^\dagger_B(t,t_0)
{\hat F_\alpha}{\hat u}_B(t,t_0),
\label{fforcela}
\end{eqnarray}
with 
\begin{eqnarray}
{\hat F_\alpha}\equiv\frac{\partial V_c(q(t),{\hat x})}{\partial q_\alpha}-
Tr(\frac{\partial V_c(q(t),{\hat x})}{\partial q_\alpha}{\hat \rho}_G(t)).
\label{fforcelb}
\end{eqnarray}
In Eq.~(\ref{fforcelb}), ${\hat \rho}_G(t)$ is the density operator 
of the total system in the Galilei transformed coordinate system. 
The ${\hat u}_B(t,t_0)$ is the time evolution operator of the subspace B, 
which satisfies the partial differential equation  
\begin{eqnarray}
i\hbar \frac{\partial}{\partial t}{\hat u}_B(t,t_0)={\hat h}_B(t)
{\hat u}_B(t,t_0),
\label{evolution}
\end{eqnarray}
with the initial condition
${\hat u}_B(t_0,t_0)=1.$
In Eq.~(\ref{evolution}), the effective Hamiltonian is given by
\begin{eqnarray}
& &{\hat h}_B(t)={\hat H}_B({\hat x})+V_c(q(t),{\hat x}),
\label{ehamilB} 
\end{eqnarray}
where ${\hat H}_B({\hat x})$ is the unperturbed Hamiltonian of space B and 
$V_c$ the coupling Hamiltonian.
The density operator ${\hat D}(t)$ describing our basic 
equations [(\ref{QFPE}), (\ref{rfunct}) and (\ref{cfunct})] 
is defined from the density 
operator in the Galilei transformed coordinate space 
${\hat\rho}_G(t)$ by 
\begin{eqnarray}
{\hat \rho}_G(t)={\hat u}_B(t,t_0){\hat D}(t){\hat u}_B^\dagger(t,t_0).
\label{ddensity}
\end{eqnarray}
The ${\hat D}_B(t)$ is defined by ${\hat D}_B(t)=Tr_A({\hat D}(t))$.

\section{Quantum effects on the fluctuation-dissipation theorem} 

The time evolution of the subspace B should in principle be determined by 
solving Eq.~(\ref{evolution}) or the corresponding reduced von Neumann 
equation \cite{tak81,niita83}. 
Here we approximate the density operator by the canonical distribution, 
\begin{eqnarray}
{\hat \rho}_B(t)\approx \exp\{\beta(t)[{\cal F}-{\hat h}_B(t)]\}.
\label{cano} 
\end{eqnarray}
We further replace ${\hat h}_B(t)$ by ${\hat H}_B$ in Eq.~(\ref{cano})
to be consistent with the linear response theory. 
One can then easily show by introducing spectral function \cite{kubo,tak81}
that the following well-known generalized fluctuation-dissipation 
theorem follows in the limit of Markovian approximation 
for the motion along a potential well,
\begin{eqnarray}
\frac{\chi^{(+E)}_{\alpha \alpha}(t)}{\chi^{(-O)}_{\alpha \alpha}(t)}
=MT^*,
\label{FDTP} 
\end{eqnarray}
with the effective temperature given by 
\begin{eqnarray}
T^*&=&\frac{1}{2}\hbar \Omega \coth\left[\frac{1}{2}\beta(t)\hbar 
\Omega\right]
\label{ETa} \\
&=&\left\{
\begin{array}{cc}
T & (T\gg \hbar \Omega) \\
\frac{1}{2}\hbar \Omega  & (T\ll \hbar \Omega).
\end{array}\right. 
\label{ETb} 
\end{eqnarray}
Equation (\ref{QFPE}) is then nothing but the Kramers diffusion equation 
\cite{kramers} postulated in Refs.~\cite{fddabe,ohta}, though 
there exist some modifications such as the temperature 
being replaced by the effective temperature.

As declared in the introduction, our interest 
in connection with the synthesis of superheavy 
elements is the diffusion process along a potential barrier instead of 
along a potential well. In this case, one needs to specify a model 
in order to further discuss the properties of the diffusion coefficients. 
We assume the Feynmann-Vernon model \cite{fv}, which has been used also by 
Caldeira and Leggett \cite{cl} to discuss macroscopic quantum tunneling, 
assume the Ohmic dissipation, and use the Drude regularization \cite{weiss} 
by introducing the following cutoff function for the spectral density 
of the environment, i.e., the sub-space B:
\begin{eqnarray}
g(\omega)=\frac{1}{1+(\frac{\omega}{\omega_c})^2}
\label{cutofffg2}. 
\end{eqnarray}

We define
\begin{eqnarray}
Y(t)\equiv\frac{\chi^{(+E)}(t)}{\chi_{\infty}^{(-O)}}\frac{1}{M},
\label{QFDTB} 
\end{eqnarray}
where 
$\chi_{\infty}^{(-O)}$
is the expression which the odd-moment of the response function takes 
in the limit of the Markovian approximation or in the asymptotic time.
The $Y(t)$ consists of three terms, two of which strongly depend on time,
\begin{eqnarray}
Y_1(t)&=&-\frac{\hbar}{4}\omega^2_c
e^{-\omega_c(t-t_0)} 
\cot\left[\frac{1}{2}\beta(t)\hbar \omega_c\right]\nonumber \\
&&{}\times \left\{\frac{1}{\omega_c-\Omega}e^{\Omega(t-t_0)}+
\frac{1}{\omega_c+\Omega}e^{-\Omega(t-t_0)}\right\},
\label{flucNd1} \\
Y_2(t)&=&\frac{1}{\beta(t)}
\sum_{n=1,2,...}e^{-\frac{\pi 2n}{\hbar \beta(t)}(t-t_0)}
\frac{\pi 2n}{\hbar \beta(t)}\frac{\omega^2_c}
{\omega^2_c-(\frac{\pi 2n}{\hbar \beta(t)})^2} 
\nonumber \\
&~&{}\times
\left\{\frac{1}{\frac{\pi 2n}{\hbar \beta(t)}-\Omega}e^{\Omega(t-t_0)}+
\frac{1}{\frac{\pi 2n}{\hbar \beta(t)}+\Omega}e^{-\Omega(t-t_0)}
\right\}.\nonumber \\
\label{flucNd2} 
\end{eqnarray}
The third one, which gives the asymptotic value, reads
\begin{eqnarray}
Y_3(t)=\frac{\omega_c^2}{\omega_c^2-\Omega^2}
\frac{1}{2}\hbar \Omega \cot\left[\frac{1}{2}\beta(t)\hbar \Omega\right].
\label{QFDTB3} 
\end{eqnarray}
Note that the right-hand side (r.h.s.) of Eq.~(\ref{QFDTB3}) 
is the analytic continuation of 
the r.h.s. of Eq.~(\ref{ETa}) concerning the sign of the barrier 
curvature if one ignores the minor change due to the first factor. 
Such analytic continuation formula of the r.h.s. of 
Eq.~(\ref{ETa}) has been argued in Ref.~\cite{hh97}. 
Our theory contains additionally non-Markovian effects. 
We note that $Y(t)$ reduces to the temperature 
at very high temperatures.

\begin{figure}
\begin{center}
\includegraphics[width=7.5cm,clip]{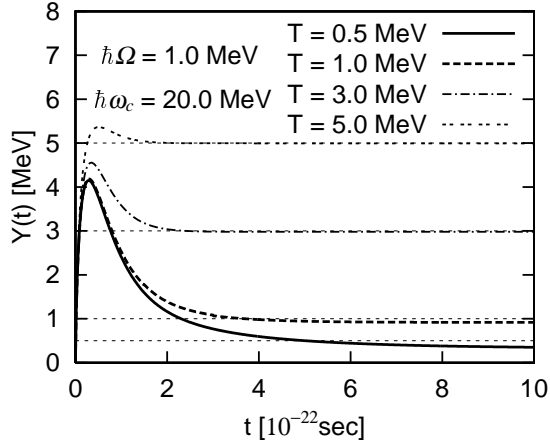}
\caption{\label{qfdt}Ratio of the diffusion to friction coefficients.}
\end{center}
\end{figure}

Figure~\ref{qfdt} shows the time dependence of $Y(t)$, i.e., 
the ratio of the diffusion to the asymptotic friction coefficients, 
for four different 
values of the temperature. The barrier curvature $\Omega $ and 
the cutoff frequency $\omega_c$ in the spectral density in the 
Drude regularization have been fixed to be $\hbar \Omega=$ 1.0 MeV 
and $\hbar \omega_c=$ 20.0 MeV. The horizontal lines 
show the position of each temperature. 
The figure shows that 
the classical fluctuation-dissipation theorem postulated 
in the original Kramers paper \cite{kramers} 
and that also in Refs.~\cite{fddabe,ohta} 
holds only at very high temperatures. The more striking result is 
the non-Markovian effect, which appears as the strong variation of 
$Y(t)$ with time. We remark that the span where $Y(t)$ shows this 
strong variation gets longer roughly in proportional to the inverse of the 
temperature as long as $T<\hbar \omega_c$.

\section{Barrier crossing probability}

We now apply our formalism to discuss quantum effects 
in the diffusion process from the fusion barrier to 
the conditional saddle in the synthesis of superheavy elements. 

We first note that the average values of $q$ and $p$ are zero in the Galilei 
transformed space, and that the solution of Eq.~(\ref{QFPE}) is a 
Gaussian. Therefore, one can set
\begin{eqnarray}
D_{AW}(p,q,t)&=&
\frac{1}{2\pi\Delta^{\frac{1}{2}}}
\exp\left[-\frac{1}{2\Delta}\sum_{i,j}y_iy_j{\tilde \sigma}_{i,j}
\right],
\label{wdfd} \\
\Delta&=&\sigma_{qq}\sigma_{pp}-\sigma_{qp}^2,
\label{msflu} 
\end{eqnarray}
where $y_1=q$ and $y_2=p$ and the $2\times 2$ matrix ${\tilde \sigma}$ is 
the inverse matrix of the $2\times 2$ matrix 
\begin{eqnarray}
\left(
\begin{array}{@{\,}ll@{\,}}
\sigma_{qq} & \sigma_{qp} \\
\sigma_{qp} & \sigma_{pp}
\end{array}
\right)
\end{eqnarray}
which determines the fluctuations, i.e., the mean square deviations from the 
average values. 
The values of $\sigma_{ij}$ are obtained by solving the 
following coupled equations given by Eq.~(\ref{QFPE}):
\begin{eqnarray}
&&\frac{d}{dt}
\left(
\begin{array}{@{\,}l@{\,}}
\sigma_{qq}(t) \\
\sigma_{qp}(t) \\
\sigma_{pp}(t) 
\end{array}
\right)\nonumber\\
&=&\!\!
\left(
\begin{array}{@{\!}ccc@{\!}}
0 &\!\! \frac{2}{M} &\! 0 \\
-(C-\chi^{(-E)}) &\!\! -\chi^{(-O)} &\! \frac{1}{M} \\
0 &\!\! -2(C-\chi^{(-E)}) &\! -2\chi^{(-O)}
\end{array}
\right)\hspace{-0.2cm}
\left(
\begin{array}{@{}l@{}}
\sigma_{qq}(t) \\
\sigma_{qp}(t) \\
\sigma_{pp}(t) 
\end{array}
\right)\nonumber\\
&&{}+
\left(
\begin{array}{@{\,}c@{\,}}
0 \\
\chi^{(+O)} \\
2\chi^{(+E)} 
\end{array}
\right)
\label{ceqflu} 
\end{eqnarray}
The Wigner distribution function for the 
macroscopic motion in the original space fixed frame 
is given by
\begin{eqnarray}
\rho_{AW}(q,p,t)=D_{AW}(q-q(t),p-p(t),t),
\label{wdfos}
\end{eqnarray}
once the Wigner distribution function in the Galilei transformed space 
$D_{AW}$ is obtained.

We represent the conditional saddle by a parabola,
\begin{eqnarray}
V_{\mathrm{cs}}(q)=-\frac{1}{2}M\Omega^2q^2,
\label{fbarrier}
\end{eqnarray}
and calculate the probability to cross the conditional saddle in 
order to form a compound nucleus by 
\begin{eqnarray}
P(t)&=&\int_0^{\infty}dq  \frac{1}{\sqrt{2\pi\sigma_{qq}(t)}}
\exp\left(-\frac{[q-q(t)]^2}{2\sigma_{qq}(t)}\right)\nonumber\\
&=&\frac{1}{2}\textrm{erfc}\left(-\frac{q(t)}
{\sqrt{2\sigma_{qq}(t)}}\right).
\label{qdprob}
\end{eqnarray}

We ignore the radial dependence of the friction tensor.
Denoting the initial position and momentum 
of the classical trajectory 
of the macroscopic variable as 
($q_0$, $p_0$), the position at time $t$ is given by
\begin{equation}
q(t)=e^{-\frac{\beta}{2}t}\!\left[q_0\!\left(\!\cosh{\frac{\beta'}{2}t}+
\frac{\beta}{\beta'}\sinh{\frac{\beta'}{2}t}\!\right)
+2\frac{p_0}{\beta'}\sinh{\frac{\beta'}{2}t}\right]\!,
\label{lantraj}
\end{equation}
with $\beta'=\sqrt{\beta^2+4\Omega^2}$.
We adopt the value of the reduced friction 
parameter $\beta$ 
from previous studies of fluctuation-dissipation 
dynamics using Langevin equation \cite{fddabe} as
$\beta=5\times 10^{21}s^{-1}$.
The curvature of the potential barrier is assumed to be 
$\hbar \Omega=1$ MeV, which is relevant to heavy nuclei.
The initial position $q_0$ is chosen to make the height of the conditional 
saddle be 4.0 MeV, and the mass parameter to correspond to the reduced mass 
in the collision of the mass number 48 and 238 nuclei. 
We defer the study of the effects of 
purely non-Markovian terms $\chi^{(-E)}$ and $\chi^{(+O)}$
and leave them in determining the $\sigma_{qq}(t)$ in the 
following analyses. 

\begin{figure}
\begin{center}
\includegraphics[width=7.5cm,clip]{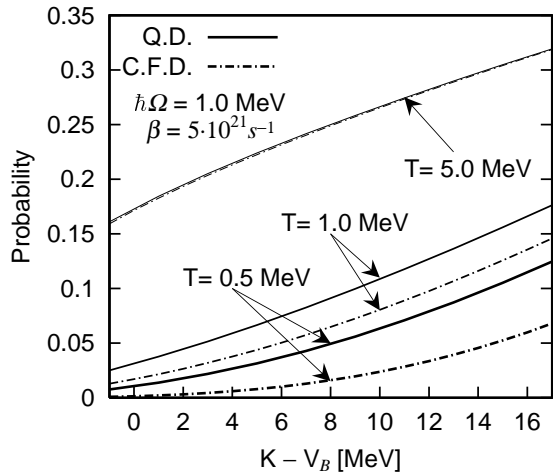}
\caption{\label{formprob}Comparison of the probability 
to cross the conditional saddle 
calculated by quantum diffusion equation and by assuming the 
classical fluctuation-dissipation theorem.}
\end{center}
\end{figure}

Figure \ref{formprob} shows the probability to cross the conditional saddle 
as a function of the initial kinetic energy $K$, 
which is measured relative to 
the height of the conditional saddle V$_B$. 
We remark that the ratio $q(t)/\sqrt{2\sigma_{qq}(t)}$ 
in Eq.~(\ref{qdprob}) converges 
to an asymptotic value. It was used to evaluate the 
probability to cross the conditional saddle $P$. 
We choose three values for the temperature.
The solid lines are the results of our theory, while the dot-dashed 
lines are the results when the classical fluctuation-dissipation 
theorem has been assumed by ignoring the quantum effects due to the 
finite curvature of the conditional saddle. The figure clearly shows that 
the quantum effect is important at low temperatures, which are 
relevant to the synthesis of superheavy elements.

\begin{figure}
\begin{center}
\includegraphics[width=8cm,clip]{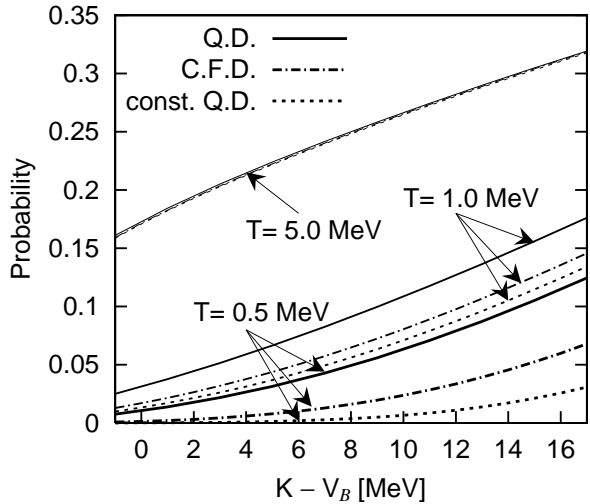}
\caption{\label{memory}Analysis of the memory effects.}
\end{center}
\end{figure}

Our theory contains a memory effect. In order to discuss 
the connection to a previous work \cite{hh97}, 
we artificially isolate the memory effect by calculating the 
probability to cross the conditional saddle 
by using the asymptotic value of the 
diffusion coefficient. The result is added in 
Fig.~\ref{memory} by the dotted line.
We observe that the probability to cross 
the conditional saddle is reduced by the quantum effect if one ignores the 
memory effect. This is because 
the asymptotic diffusion coefficient in the 
quantum theory is smaller than that obtained from 
the classical fluctuation-dissipation theorem. 
A similar effect has been shown in Ref.~\cite{rh03}. 
Our study shows in addition that 
the memory effect overcomes this effect and finally the net 
quantum effects enhance the transmission probability 
of the conditional saddle.
In other words, the net quantum effects reduce the fusion hindrance.

In passing, we wish to mention that 
quantum effects on diffusion process are also discussed in 
Ref.~\cite{bb02} following a different approach. 
However, there are some important differences 
in the expressions of transport coefficients. For example, 
the diffusion coefficient given by Eq.(8) in Ref.~\cite{bb02} does 
not seem to match with our asymptotic formula (23), as well as 
those presented in Refs.~\cite{hh97,rh03}. We also wish to refer to 
Ref.~\cite{my00}, which discusses the dynamics of barrier penetration 
in a thermal medium for the inverted harmonic oscillator by using 
the influence functional formalism of the path integral method.

\section{Summary and future developments}

We have presented a diffusion theory which takes the finite 
curvature of the potential field into account. The theory is then 
applied to the case where the temperature, 
barrier curvature, and the friction 
coefficient are taken to represent a realistic 
situation of the diffusion process from the fusion barrier to 
the conditional saddle in the synthesis of superheavy elements. 
We have thus shown that the quantum effects will play an important 
role. 
We especially pointed out the importance of the memory effect 
which has been omitted in any previous works. It makes the 
net quantum effects enhance the probability of crossing 
the conditional saddle, 
while the quantum effect reduces it if the memory effect is ignored, 
as has been shown in Ref. \cite{rh03}.  

We have artificially left out some of the genuine non-Markovian 
terms, i.e., the odd moment of the correlation function and the 
even moment of the response function. Also, we have assumed a sharp 
distribution at the initial time and left out the effects of 
spreading of the initial distribution. We will discuss these effects  
as well as the dependence of the quantum effects on 
the strength of the dissipative force in forthcoming papers.  
One of the interesting problems is to clarify whether our conclusion 
concerning the role of non-Markovian effect is special to our 
specific choice of the Caldeira-Leggett model, 
especially to the Ohmic dissipation, or holds in general.
This is another issue which we will explore in the near future.

\section*{ACKNOWLEDGMENTS}
This work is supported in part by
the Grant-in-Aid for Scientific Research from 
Ministry of Education, Culture, Sports, Science and Technology 
under Grant No. 12047203 and No. 13640253, 
also under the Special Area Research, Contract No. 08640380, 
and also by the U.S. DOE Grant No.DE-FG05-89ER40530. 
N.T. thanks Tennessee Technological University for support 
and hospitality during his visits. 
N.T. and S.A. thank also A. Gokalp and O. Yilmaz  for their 
kind hospitality at the Middle East Technical University/Ankara, 
where this work has been partly completed.

\end{document}